\documentclass{iaus}
\usepackage{graphicx}

\title[Spectropolarimetry of SNR G296.5+10.0] 
{Spectropolarimetry of supernova remnant G296.5+10.0}

\author[Harvey-Smith L., Gaensler B.M., Ng C.-Y. \& Green A. J.]   
{Lisa Harvey-Smith, Bryan M. Gaensler, C.-Y. Ng \break \and Anne J. Green}

\affiliation{Sydney Institute for Astronomy (SIFA), School of Physics,\break The University of Sydney, NSW 2006, Australia.\break email: lhs@usyd.edu.au\\[\affilskip]}

\pubyear{2009}
\volume{259}  
\pagerange{100--100}
\date{Dec. 1st 2008 and in revised form ??}
\jname{Cosmic Magnetic Fields: From Planets, to Stars and Galaxies}
\editors{K.G. Strassmeier, A.G. Kosovichev \& J.E. Beckman, eds.}
\begin{document}

\maketitle

\begin{abstract}
Radio continuum emission from the supernova remnant G296.5+10.0 was observed using the Australia Telescope Compact Array. Using a 104 MHz bandwidth split into 13 $\times$ 8 MHz spectral channels, it was possible to produce a pixel-by-pixel image of Rotation Measure (RM) across the entire remnant. A lack of correlation between RM and X-ray surface brightness reveals that the RMs originate from outside the remnant. Using this information, we will characterise the smooth component of the magnetic field within the supernova remnant and attempt to probe the magneto-ionic structure and turbulent scale sizes in the ISM and galactic halo along the line-of-sight.
\keywords{Polarization -- Magnetic fields -- (ISM:) supernova remnants, radio continuum: ISM, X-rays: ISM}
\end{abstract}

\firstsection

\begin{figure}
\begin{center}
\includegraphics[width=60mm, angle=-90]{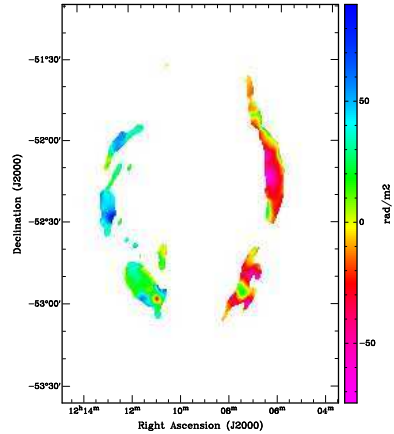}
\caption{Rotation Measure image of G296.5+10.0}
\end{center}
\end{figure}

\section{Observations}
The supernova remnant G296.5+10.0 was observed using the Australia
Telescope Compact Array at 1.4 GHz with an angular resolution of 30 arcseconds.
Signals from the dual linear polarized feeds were combined to produce images of
the remnant in linear polarisation. The Rotation Measure synthesis method (Brentjens \& de Bruyn, 2005) was employed to obtain an RM spectrum at each pixel in the image (Figure 1). The 104 MHz bandwidth was split into 13 $\times$ 8 MHz spectral channels, giving sufficient coverage in $\lambda$$^2$ to remove any n$\pi$ ambiguities in the RM values. X-ray data in the energy range 0.1--2.4 keV were taken from the ROSAT archive, via NASA's SkyView. These complimentary data allow us to estimate the electron density in the remnant.

\section{Determination of n$_e$ and $\bf{B}$$_{||}$}

In order to disentangle the electron density (n$_e$) and line-of-sight magnetic field ($\bf{B}$$_{||}$) we measured the X-ray intensity at every point in the ROSAT image, after regridding and smoothing them to match the ATCA image. 
RM is defined as the integral of n$_e$ and $\bf{B}$$_{||}$ along the line of sight. The thermal X-ray surface brightness, S$_x$, is proportional to n$_e$$^2$. Therefore if RM and S$_x$$^{1/2}$ are correlated, the Faraday rotation is occurring within the thermal plasma (Matsui et al. 1984).
A plot of  RM vs. S$_x$$^{1/2}$  (Figure 2) shows that there is no clear correlation between RM and S$_x$$^{1/2}$ in the remnant, which means that the RMs originate from the ISM/halo along the line-of-sight. Therefore the RM variations across the remnant may reflect the structure of the magnetised interstellar medium (Haverkorn et al. 2004) and the smooth component may tell us about the overall magnetic field morphology within the remnant (Ransom et al. 2008). This will form the basis of our further analysis of these data.

\begin{figure}
\begin{center}
\includegraphics[width=65mm, angle=-90]{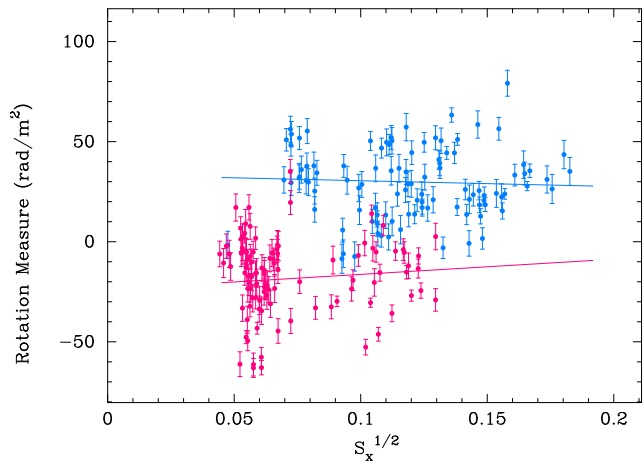}
\end{center}
\caption{Plots of (X-ray surface brightness)$^{1/2}$ against Rotation Measure for the east (blue) and west (pink) sides of the remnant. A least-squares best fit is plotted onto each data set.}
\end{figure}

\section{On-going work}\label{sec:concl}
The following analytical steps will be taken:
(1) Modelling of the RM structure across the remnant to determine the large-scale structure and morphology of the magnetic field (i.e. linear, spiral, toroidal) in the remnant (2) Analysis of smaller-scale RM variations across the remnant, to probe the turbulent scale-sizes in the ISM and galactic halo.

\begin{acknowledgments}
Lisa Harvey-Smith acknowledges the IAU for a travel grant for this symposium.
\end{acknowledgments}

\end{document}